\documentclass[submission,11pt%
 ]{dmtcs}
\usepackage[latin1]{inputenc}
\usepackage{graphicx}
\usepackage{amsmath,amsfonts,amssymb}

\usepackage[round,sort]{natbib}       
\bibpunct{[}{]}{,}{n}{}{;}     

 \usepackage{setspace}

\usepackage{verbatim}
\usepackage{graphicx}
\usepackage{epsf}
\usepackage{epsfig}

\makeatletter

\DeclareMathSymbol{\Gamma}{\mathalpha}{letters}{"00}
\DeclareMathSymbol{\Delta}{\mathalpha}{letters}{"01}
\DeclareMathSymbol{\Theta}{\mathalpha}{letters}{"02}
\DeclareMathSymbol{\Lambda}{\mathalpha}{letters}{"03}
\DeclareMathSymbol{\Xi}{\mathalpha}{letters}{"04}
\DeclareMathSymbol{\Pi}{\mathalpha}{letters}{"05}
\DeclareMathSymbol{\Sigma}{\mathalpha}{letters}{"06}
\DeclareMathSymbol{\Upsilon}{\mathalpha}{letters}{"07}
\DeclareMathSymbol{\Phi}{\mathalpha}{letters}{"08}
\DeclareMathSymbol{\Psi}{\mathalpha}{letters}{"09}
\DeclareMathSymbol{\Omega}{\mathalpha}{letters}{"0A}

\def\bi{\begin{itemize}}
\def\ei{\end{itemize}}
\def\indexname{Index}
\makeatother

\usepackage{verbatim}
\usepackage{graphicx}   

\newcommand{\bq}{\begin{eqnarray*}}
\newcommand{\eq}{\end{eqnarray*}}

\def\nmb{\mbox{\#ones}}

\newcommand{\be}{\begin{equation}}
\newcommand{\ee}{\end{equation}}

\newcommand{{\integer}}{\mbox{\rm Z\hspace{-0.85em}Z}\,}

\def\squareforqed{\hbox{\rlap{$\sqcap$}$\sqcup$}}
\def\qed{\ifmmode\squareforqed\else{\unskip\nobreak\hfil
\penalty50\hskip1em\null\nobreak\hfil\squareforqed
\parfillskip=0pt\finalhyphendemerits=0\endgraf}\fi}

\def\blacksquareforqed{\hbox{$\blacksquare$}}
\def\bqed{\ifmmode\blacksquareforqed\else{\unskip\nobreak\hfil
\penalty50\hskip1em\null\nobreak\hfil\blacksquareforqed
\parfillskip=0pt\finalhyphendemerits=0\endgraf}\fi}

\begin{document}
\bibliographystyle{plainnat}

\author{Joel Ratsaby\thanks{\url{ratsaby@ariel.ac.il}{\ }}
}
\title[]{An algorithmic complexity interpretation of Lin's third law of information theory}
\address{
Electrical and Electronic Engineering Department\\
Ariel University Center of Samaria\\
Ariel 40700, ISRAEL\\
}

\keywords{Entropy, Randomness, Information theory, Algorithmic complexity, Binary sequences}

\revision{1.0}
\received{\today}

\maketitle

\begin{abstract}
Instead of static entropy we assert  that the Kolmogorov complexity of a static structure such as a solid
is the proper measure of disorder (or chaoticity).
A static structure   in a surrounding perfectly-random universe 
 acts as an interfering entity which introduces local disruption in  randomness.
This is modeled by a selection rule $R$ which selects a subsequence of the random input sequence that
hits the structure.
Through the inequality that relates
 stochasticity and chaoticity of random binary sequences 
 we maintain that Lin's notion of stability corresponds to the stability of the frequency
 of $1$s in the selected subsequence. This explains why more complex static structures are less stable.
 Lin's third law is  represented as the inevitable change that static structure undergo towards
  conforming to the universe's perfect randomness.
\end{abstract}



\section{Introduction}

Lin \cite{SKlin1996,SKlin2001,sklin2008}
 introduced a 
new notion of entropy, called {\em static} entropy.
His aim was to revise information theory 
in order to broaden the notion of entropy such that 
the naturally occurring phenomenon of increased symmetry of a system and 
the increased similarity of its components may be explained
by information theoretic principles.

He proposed that entropy is the degree of symmetry and information is the degree of asymmetry of
a structure. With respect to the second law of thermodynamics 
 (which states that entropy of an isolated
system increases to a maximum at equilibrium)
any spontaneous (or irreversible) process evolves towards a state of maximum symmetry.
Many instances of spontaneous processes of different types
exist in nature: molecular interaction, phase
separation and phase transition,  symmetry breaking and the densest molecular
packing, crystallization,
self-aggregation, self-organization,
chemical processes
 used to separate or purify different substances.
These are just a few of the many examples of processes which, according to Lin,
are inherently driven by an information minimization or symmetry
maximization process.

Spontaneous processes  lead to
 products 
 that are  simpler and have a more regular (more symmetric) structure
 of a  lower information content. 
As Lin \cite{SKlin2001} suggests, in a heterogenous structure
consisting of a rich variety of components,  temperature plays a key role
in controlling the direction of a process.
Increasing it leads to  a homogeneous mixing of the components  while 
decreasing it leads to a spontaneous phase separation of the components that differ.
The latter leads to a system of a lesser information content.

We assert that Lin's notion of information with regards to such processes
is better represented 
by the information needed to describe an object (algorithmic information)
rather than Shannon's notion of information and entropy
which is applicable to settings underlaid by  stochastic laws.
For instance, a simple 
static structure such as a crystal can be described by a short binary string hence
has little amount of information.
As Lin \cite{SKlin2001} suggests,
 symmetric static structures (crystals)
and nonsymmetrical static structures
have different amounts of descriptive information.

Lin realizes the incompatibility of Shannon's information theory 
in explaining  spontaneous processes 
and suggests that the classical definition of entropy as a measure of 
 disorder has been  incorrectly used in the associated fields of science.
Clearly, a static structure (not necessarily a crystal), which is a frozen dynamic structure,
is in a more orderly state than a system of dynamic motion.
But when comparing two static structures
the  symmetric one should be measured as having more
order  (less chaotic),  being simpler and having less information content
than the  nonsymmetrical structure. 
As he remarks, this is not explained well
by the classical (Shannon) definitions of entropy and information.
According to these definitions,
 a more symmetric and  ordered system is
  obtained 
 by the introduction of additional information, i.e., by
a reduction in  entropy. 
This seems to contradict
 observed spontaneous processes (such as those mentioned above)
 which occur automatically without any external (information agent) intervention while still
 leading to the  formation of
  simpler  and more stable systems that have a higher degree of symmetry 
(and less information-content).

Lin's approach to explaining this
is to  generalize
the notion of entropy (although, this is not defined precisely) also for static (deterministic) objects, i.e., nonprobabilistic structures.
He then postulates
a third law  of information theory which states that 
maximal static entropy (and zero information content)
 exists  in a perfect crystal at zero absolute temperature.
Together with  the second law of information theory (which parallels the second
law of thermodynamics) this forms
the basis for explaining  the above processes.



Lin has made  important and deep contribution in making the connection
between the concepts of entropy, simplicity, symmetry and similarity, but
as he writes (on p. 373, \cite{SKlin1996}), "..it is more important to 
 give an information content  formula  for static structures" and then
 remarks that to his knowledge there is no such formula. 

In this paper we  develop such formulation and introduce an alternate explanation
to his notion of static entropy.
Our approach is based on the area of algorithmic information theory
which relates concepts of randomness and complexity of finite objects
(since any finite object may be described by a
finite  binary sequence we henceforth refer to them simply
as binary sequences).

We explain Lin's principle of higher entropy---higher symmetry
through the mathematical constructs of algorithmic information theory.
We  state a  precise quantitative definition
of the  information content of an object (what Lin calls a static structure or solid) and then give statements
about the relation between
algorithmic complexity (this replaces Lin's notion static entropy),
information content
(of an object such as a solid)
 and its stability (which Lin associates with symmetry).

Starting in the next section
we introduce several concepts from this area.

\section{Algorithmic complexity}

Kolmogorov \cite{Kolmogorov65} 
 proposed to measure the conditional complexity of a finite object $x$ given 
a finite object $y$ by the length of the shortest binary sequence $p$ (a program for computing
$x$) which consists of $0$s and $1$s and which reconstructs $x$ given $y$.
Formally, this is defined as 
\be
\label{kxy}
K(x|y) = \min\{l(p) : \phi(p,y) = x\}
\ee
where $l(p)$ is the length of the sequence $p$, $\phi$ is a
universal partial recursive function which acts as a description method, i.e.,
when provided with input $(p, y)$ it gives 
a specification for $x$. 
The word universal means that the function $\phi$ can
 emulate any Turing machine (hence any partial recursive function).
One can view $\phi$ as a universal computer that can
interpret any programming language and accept any valid program $p$.
The Kolmogorov complexity  of $x$ given $y$ as defined in (\ref{kxy})  is the length of the shortest program
that generates $x$ on this computer given $y$  as input.
The special case of $y$ being the empty binary sequence gives the unconditional
Kolmogorov complexity $K(x)$.
This has been later extended by \cite{Gac74,Chaitin75}
 to the prefix-complexity which requires $p$ to be coded in a prefix-free format.

 \section{Algorithmic randomness}
 
  The notion of randomness or stochasticity of finite objects (binary sequences)
aims to explain  the intuitively clear notion  that a
   sequence, whether finite or infinite, should be measured as being more unpredictable
   if it possess fewer regularities (patterns).
    This conceptualization underlies a fundamental notion 
    in information theory and in the area known as algorithmic information theory
    which states that
    the existence of regularities underlying data 
    necessarily implies information redundancy in the data (observations).
This notion has become the basis of the well-known principle of  
 Minimum Description Length (MDL)
 \citep{Rissanen89} which was proposed as a
     computable approximation of Kolmogorov complexity and has since served
     as a powerful method for inductive inference. 
     The MDL principle states that 
     given a limited set of observed data
     the best explanation or  model (hypothesis)
       is the one 
     that permits the greatest compression of the data. 
     That is, the more we are able to compress the data, the more we learn about the underlying regularities that generated the data.

The area  known as {\em algorithmic randomness} 
studies the relationship between complexity and stochasticity (the degree of unpredictability) of
 finite and infinite binary sequences \citep{LaurentBienvenu07}.
Algorithmic randomness was first considered  by  von Mises in 1919 who
defined an infinite
binary sequence $\alpha$ of zeros and ones as `random' if it is unbiased, i.e. if the
frequency of zeros goes to $1/2$, 
and  every subsequence of  $\alpha$
that we can extract
using  an `admissible' selection rule (see 
\S4)
 is also not biased. 

Von Mises never made completely precise what he meant
by admissible selection rule.
Church proposed a formal definition of such a rule to
 be a (total) computable process which, having read the first $n$ bits of an
infinite binary sequence $\alpha$, decides if it wants to select the next bit or not, and
then reads it (it is crucial that the decision to select the bit or not
is made before reading the bit). The sequence of selected bits is the selected
subsequence with respect  to the selection rule.
Kolmogorov and Loveland \cite{Loveland66,Kolmogorov98}  
argued that
in Church's definition the bits are read in order, which is too restrictive.
They
proposed
a more permissive definition of an admissible selection rule as any  (partial) computable
process which, having read any $n$ bits of an infinite binary sequence $\alpha$, picks
a bit that has not been read yet, decides whether it should be selected or not,
and then reads it. When 
subsequences selected by such a selection rule pass the unbiasness test
they are called Kolmogorov-Loveland stochastic (KL-stochastic
for short).

It turns out that even with this improvement, KL-stochasticity is too weak a
notion of randomness. Shen \cite{Shen89}   showed that there exists a KL-stochastic sequence all of whose prefixes
contain more zeros than ones.
 Martin L$\ddot{o}$f \cite{MartinLof66} introduced a notion of randomness which
is now considered by many as the most
satisfactory notion of algorithmic randomness.
Martin-L$\ddot{o}$f's definition says
precisely which infinite binary sequences are random and which are not. The definition is probabilistically
convincing in that it requires each random sequence to pass every algorithmically implementable
statistical test of randomness.
From  the work of \cite{Shnor71} 
Martin-L$\ddot{o}$f randomness can be characterized in terms of Kolmogorov
complexity (\ref{kxy}) of $\alpha$.
An infinite binary sequence 
$\alpha = \{\alpha_i\}_{i=1}^\infty$
is Martin-L$\ddot{o}$f random if and only if there is a constant $c$
such that for all $n$,  $K(\alpha_1, \ldots, \alpha_n) \geq
n - c$ where $K$ is  the prefix Kolmogorov complexity \citep{citeulike:1877660}.


\section{Selection rule}
\label{sr}
 
In this section we describe the notion of a selection rule. As
mentioned in the previous section this is a principal concept used
as part of   tests of  randomness of sequences. In later sections we 
use this to develop a  framework 
for explaining  Lin's third law.

 An admissible {\em selection rule} $R$ is defined \citep{Vyugin99} based on three partial
 recursive functions $f, g$ and $h$ on $\Xi$.
 Let $x=x_1, \ldots, x_n$. The process of 
selection is recursive. It begins with an empty sequence $\emptyset$. The function $f$ 
is responsible for selecting possible candidate bits of $x$
as elements of the subsequence to be formed. The function $g$  examines the value of these bits
and decides whether to include them in the subsequence. Thus $f$  does so 
according to the following definition:  $f(\emptyset)=i_1$, and
 if at the current time $k$ a subsequence has already been selected which consists of elements
$x_{i_1}, \ldots, x_{i_k}$ then  $f$  computes the index of the next element to be examined according to
element $f(x_{i_1}, \ldots, x_{i_k})=i$ where $i\not\in \{i_1, \ldots, i_k\}$, i.e., 
the next element to be examined must not be one which has already been selected
(notice that maybe $i<i_j$, $1\leq j\leq k$, i.e., the selection rule
can go backwards on $x$). Next, the two-valued
function $g$ selects this element $x_i$ to be the next element of the constructed
subsequence of $x$ if and only if $g(x_{i_1}, \ldots, x_{i_k}) = 1$.
The role of the  two-valued function $h$
is to decide  when this process must be terminated.
This subsequence  selection process terminates if $h(x_{i_1}, \ldots, x_{i_k}) = 1$ or
$f(x_{i_1}, \ldots, x_{i_k}) > n$. Let $R(x)$ denote the selected subsequence.
By $K(R|n)$ we mean the length of the shortest program computing the values
of $f$, $g$ and $h$ given $n$.

\section{Randomness deficiency}

Let $\Xi$ be the space of all finite binary sequences and
denote by $\Xi_n$  the set of all finite binary sequences of length $n$.
Kolmogorov introduced a notion of {\em randomness deficiency} 
 $\delta(x|n)$ of a finite sequence $x\in \Xi_n$  as follows:
\[
\delta(x|n) = n - K(x|n)
\]
 where $K(x|n)$ is the Kolmogorov complexity of $x$
 not accounting for its length  $n$, i.e., it is a measure of complexity
  of the information that codes only the specific pattern of $0$s and $1$s in $x$ 
 without   the bits that code the length of $x$  (which is $\log n$ bits).
  Randomness deficiency measures  the opposite of chaoticity of a sequence.
  The more regular the sequence the less complex (chaotic) and the higher its
  deficiency.
  
\section{Biasness}

 From the previous sections
 we know that there are two principal measures related to the information content
 in  a finite sequence $x$,  stochasticity (unpredictability)
 and chaoticity (complexity).
An infinitely long  binary sequence
is regarded random
if it satisfies the principle of stability  
of the frequency
of $1$s for any of its subsequences
that are obtained by an admissible selection rule \citep{Kolmogorov63,Kolmogorov98}.

 Kolmogorov \cite{Kolmogorov65} showed that the stochasticity of
a finite binary sequence $x$ may be precisely expressed by the deviation of the
 frequency of ones from some $0<p<1$,
  for any subsequence of $x$ selected by an admissible selection rule $R$ 
 of finite complexity $K(R|n)$.
  The chaoticity of $x$
  is the opposite of its randomness deficiency, i.e., it is large
  if  its Kolmogorov complexity
  is close to its length $n$.
The works of
\cite{Kolmogorov65,Asarin88,Asarin87,Vyugin99}
  relate this chaoticity to stochasticity.
In \cite{Asarin88,Asarin87} it is shown that chaoticity implies stochasticity. This  can be seen from
  the following relationship (with $p=1/2$):
\begin{equation}
\label{Ineq}
\left|\nu(R(x)) - \frac{1}{2}\right| 
\leq c\sqrt{ \frac{ 
	\delta(x|n) + K(R|n) + 2\log K(R|n) }{l(R(x))}}
\end{equation}
where for a binary sequence $s$,
$\nmb(s)$ denotes the number of $1$s in $s$,
 $\nu(s) = \nmb(s)/l(s)$ represents the frequency of $1$s 
 in $s$,
     $l(R(x))$ the length of the
 subsequence selected by $R$
 and  $c>0$ is some absolute constant.

From this we see that 
as the chaoticity of $x$ grows (the randomness deficiency decreases) the stochasticity of the selected subsequence  grows, i.e.,
the bias from $1/2$ decreases.
 The  information content of the selection rule, namely $K(R|n)$,
has a direct effect on this relationship: the lower $K(R|n)$ is the stronger
the stability (i.e., the smaller the deviation of the frequency of $1$s from $1/2$).
In
  \cite{Vereshchagin}
the other direction is proved which
shows that stochasticity implies chaoticity.

\section{Interpreting a static structure  as a selection rule}

In this section we apply the above theory to explain  Lin's principle
of symmetry, static entropy and stability whose key features and relationship are
summarized again here below:

\bi

\item higher symmetry---higher static entropy  (Lin's third law of information theory)
\item higher symmetry---higher stability (the most symmetric microstate will be the most stable static structure)
\item  higher static entropy---higher loss of information (a system with high information-content is unstable, spontaneous phase separation is accompanied by an information loss)

\ei

\begin{figure}[h]
\begin{center}
\epsfxsize=3in
\includegraphics[clip=true,scale = .6]{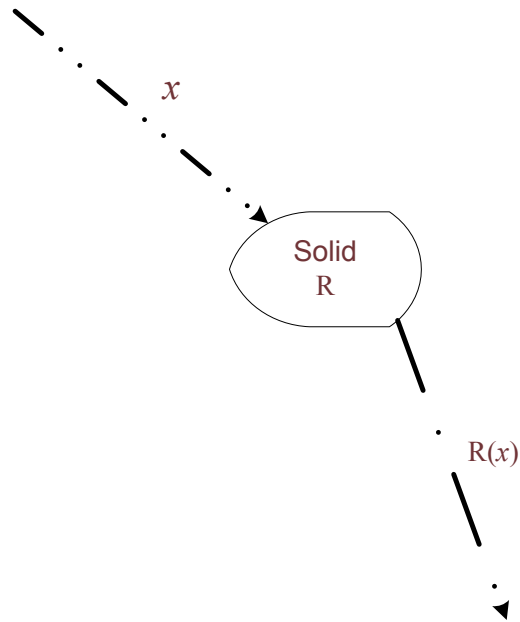}
\end{center}
\caption{Interaction of solid with surrounding space}
\label{fig1}
\end{figure}

Consider a solid
which
represents a static 
structure in which  information is registered \citep{SKlin1996}. 
The surrounding universe
consists of a dynamic system of particles randomly hitting the solid, some are deflected
and some pass through. Let us represent this as a binary sequence $x$
with $x_i=1$ representing a particle hitting the solid
and $x_i=0$ indicating the absence of a particle  at discrete time $i$.
This surrounding universe is as close to what can be considered
as perfect (or true) source of randomness. Thus the randomness deficiency of $x$ is at its minimum.
We associate with the solid 
 a selection rule $R$ which acts by realizing an algorithm of complexity $K(R)$
which selects certain bits from the $x$ and
decides whether  to let the bit be in the output subsequence
 $R(x)$ of $x$ (Figure \ref{fig1}).
Note that  the formed subsequence $R(x)$ 
may consist of not only $1$s but also $0$s since
discrete time $i$ (which indexes the bits)
 keeps increasing also when no  particle hits the solid and hence when 
 no particle
emerges from the back surface of the solid.
 
 The selection mechanism of particular bits of the input 
 sequence $x$ by the solid (similar to the way $f$ 
 is defined for a selection rule in \S4)
 is due to the solid's complex response to the presence 
 of a sequence of particles hitting it.
 For instance, a solid's dielectric constants or material 
 nonhomogeneity cause a nonlinear 
 response to particle bombardment, e.g.,
 photoconductive solid materials bombarded by photon 
 particles \cite{odagawa:016101} and
 superconductive thin films hit by ions \cite{0295-5075-6-5-009}
  exhibit transient responses. This is analogous to the
 impulse response of a non-perfect filter in communication systems.
 Materials science applications involving pulsed beams show that the
time dependent response of a solid to a high energy ion beam is 
described by complex 
 hydrodynamic codes \cite{energy1}. 
  Consider a random electronic signal consisting of a sequence $x$ of binary 
 pulses which inputs to an electronic switching
 device with a transient response governed by  time constants. 
 Then for certain segments  of $x$ where the
pulses frequency is high the device will not manage to track exactly every bit change
 in the input sequence $x$ 
 and as a result its output will correspond only
 to a certain subsequence of $x$.
 This inability to track the input sequence in a precise
 way implies that the device (and similarly a solid acting on a random
 sequence of particles) effectively selects certain bits from $x$ and `decides'
 which to pass through.


A simple solid is one whose information content is small. 
Its selection behavior is of low
complexity $K(R)$ since it can be described by a more concise 
time-response model (shorter `program').
We assert that
representing the complexity of a static structure by
the Kolmogorov complexity of its corresponding selection rule $R$
allows to make Lin's qualitative notion of static entropy 
more robust.
  Lin's notion of stability
can alternatively  be represented  by  the stability of
 the frequency of $1$s in the selected subsequence $R(x)$.
Provided that  $x$ has a small randomness deficiency $\delta(x|n)$
it follows from  (\ref{Ineq}) 
 that
 for a solid  with  low information content (small $K(R|n)$)
the deviation of the frequency of $1$s from $1/2$ will be small.
This means that the frequency stability is large.
We assert that this notion of stability corresponds with Lin's notion of physical
stability of the system, i.e., the stability of a static structure.

A more  complex static structure (solid $R$ with large $K(R|n)$) leads to a higher deviation, i.e., lower stability, of the frequency
of $1$s from $1/2$ in the selected subsequence $R(x)$. We assert that this is in correspondence  with Lin's notion
of instability of a system with high information content.

 Lin's third law states that any spontaneous process (irreversible) 
tends to produce  a structure of a higher static entropy which in his words "minimizes information and maximizes symmetry".
According to our framework, this third law is explained as follows:
{\em over the course of time (and provided there are no external intervention or constraints) 
a solid structure will eventually not be able to  maintain its `interference'
with the perfect randomness of its surrounding universe
(viewed here as a selection rule, interference means that the selected subsequence has a frequency of $1$s
which is biased away from $1/2$).
At some point, the static structure cannot maintain its resistance to randomness
and  becomes algorithmically less complex and thereby 
 moves into a more stable state.}

In summary, our explanation of Lin's third law is as follows: 
 static  structures (compared to dynamic systems) 
 are like points of interference in space of perfect randomness.
 They
 resist
 randomness of the surrounding universe and as a result are less stable.
Static structures cannot maintain their complexity indefinitely.
Eventually, they undergo  change in the direction of being less algorithmically complex
and this makes their interference with the randomness of the surrounding universe
 weaker. As a result they become more stable.

 In agreement with Lin's assertion that a perfect crystal has zero information content
 and maximal static entropy, we assert that given any random sequence $x$ of length $n$
 then the selection rule $R$ associated with 
 Lin's crystal has a small (constant with $n$)  Kolmogorov complexity $K(R|n)$
 and the  selected subsequence $R(x)$ (which is also random)
  has a maximum stability of the frequency of $1$s.

\section{Conclusions}

This paper introduces an algorithmic complexity
 framework for representing Lin's concepts  of static entropy, stability 
and their connection to the second law of thermodynamic.
Instead of static entropy we assert  that the Kolmogorov complexity of a static structure such as a solid
is the proper measure of disorder (or chaoticity).
We consider a static structure  being in a surrounding perfectly-random universe 
in which it acts as an interfering entity which introduces a local disruption of randomness.
This is modeled by a selection rule $R$ which selects a subsequence of the random input sequence that
hits the structure.
Through the inequality that relates
 stochasticity and chaoticity of random binary sequences 
 we maintain that Lin's notion of stability corresponds to the stability of the frequency
 of $1$s in the selected subsequence. This explains why more complex static structures are less stable.
 Lin's third law is  represented as the inevitable change that static structure undergo towards
  conforming to the universe's perfect randomness.




\bibliographystyle{plainnat}


\end{document}